\documentclass[aps,prb,longbibliography,showpacs,twocolumn,superscriptaddress,amsmath,amssymb]{revtex4-2}
\usepackage{amsmath}
\usepackage{epsfig}
\usepackage{chemformula}
\usepackage{graphicx}
\usepackage{natbib}
\usepackage{array}
\usepackage{setspace}
\usepackage{amssymb}
\usepackage{soul}
\usepackage[breaklinks=true,colorlinks,linkcolor=blue]{hyperref}
\usepackage{float}
\usepackage[normalem]{ulem}

\usepackage[yyyymmdd]{datetime}

\newcommand{\SPhide}[1]{{}}

\begin{document}

\title{ Effects of Ru-doping on the magnetism of 
Ag$_{3}$LiIr$_{2}$O$_{6}$, a candidate Kitaev quantum spin liquid}

\author{Sanjay Bachhar}
\thanks{\href{mailto:sanjayphysics95@gmail.com}{sanjayphysics95@gmail.com}}
\affiliation{Department of Physics, Indian Institute of Technology Bombay, Powai, Mumbai 400076, India}

\author{M. Baenitz}
\affiliation{Max Planck Institute for Chemical Physics of Solids, 01187 Dresden, Germany}

\author{John Wilkinson}
\affiliation{ISIS Pulsed Neutron and Muon Source, STFC Rutherford Appleton Laboratory,	Harwell Campus, Didcot, Oxfordshire OX110QX, UK}

\author{A.V. Mahajan}
\thanks{\href{mailto:mahajan@phy.iitb.ac.in}{mahajan@phy.iitb.ac.in}}
\affiliation{Department of Physics, Indian Institute of Technology Bombay, Powai, Mumbai 400076, India}

\date{\today}

\begin{abstract}
We report our investigations on \ch{Ag3LiIr_{1.4}Ru_{0.6}O6} which results from the Ru substitution in the Kitaev quantum spin liquid candidate \ch{Ag3LiIr2O6}. It crystallizes in the monoclinic  $C2/m$ space group like its parent compound, \ch{Ag3LiIr2O6}. Our susceptibility measurements reveal an effective moment $\mu_{\text{eff}}=$ 2.6 $\mu_{B}$ which is higher than moments of the parent compound and less than that of the  Ru-analog (\ch{Ag3LiRu2O6}) suggesting the presence of  magnetic Ir$^{4+}$ ($J_{\text{eff}}=1/2$) and Ru$^{4+}$ ($S=1$). Bulk magnetic susceptibility suggests long-range order (LRO)at $T \sim$ 20 K whereas no  clear signature is present in the heat capacity. Likewise, there is a loss of the $^{7}$Li
NMR spectral intensity around $T \sim$ 20 K as expected at the onset of LRO but a complete wipe out is not seen in contrast to the result in \ch{Ag3LiIr2O6}.
There is as well a $T \sim$ 20 K anomaly in the $^{7}$Li
NMR relaxation rate  1/$T_{1}$ and also a fall in the $^{7}$Li NMR shift $K$ with decreasing temperature. These results
suggest LRO at  $T \sim$ 20 K in \ch{Ag3LiIr_{1.4}Ru_{0.6}O6}. However, at low-$T$ below 10 K,
we observe a power law variation in magnetic heat capacity $C_{m}$ and spin lattice relaxation rate $1/T_{1}$, temperature-independent $^{7}$K and no further
loss of the $^{7}$Li NMR spectral intensity. These results might suggest the persistence or stabilisation of a  quantum spin liquid-like phase perhaps from a fraction of the sample in \ch{Ag3LiIr_{1.4}Ru_{0.6}O6} below 10 K.  Our muon spin relaxation$\mu$SR measurements suggest ordering around 20 K consistent with our other probes. It appears that the main effect of Ru-substitution is to shift the LRO to a higher temperature in comparison with \ch{Ag3LiIr2O6} though there are signatures of a novel phase below about 10 K. 
\end{abstract}
\maketitle

\section{Introduction}
\label{sec:intro}

Kitaev's groundbreaking theoretical proposal that bond-dependent magnetic interactions can stabilize a unique $Z_{2}$-spin-liquid ground state with Majorana excitations, along with the material-specific insights of Jackeli and Khaliullin \cite{Jackeli2009}, has driven significant experimental efforts to realize such systems. Their work emphasized that honeycomb lattice structures composed of $4d/5d$ transition metal oxides, featuring edge-sharing oxygen octahedra and strong spin-orbit coupling, provide the essential ingredients for realizing the Kitaev model \cite{Kitaev2006}. Several potential candidates with such layered honeycomb structures have been explored, including \ch{Na2IrO3} \cite{Khaliullin2013}, $\alpha$-\ch{Li2IrO3} (along with its three-dimensional polymorphs) \cite{Coldea2016}, and $\alpha$-\ch{RuCl3} \cite{Do2017}. However, these materials exhibit magnetic ordering \cite{Hou2017} due to the influence of Heisenberg and other non-Kitaev interactions, limiting the observation of Kitaev interactions to either higher temperatures or when an external magnetic field is applied. An addition to this family is \ch{H3LiIr2O6} \cite{Bette2017}, which is considered as promising spin-orbit-entangled quantum spin liquid \cite{Kitagawa2018}. While \ch{H3LiIr2O6} exhibits no magnetic ordering down to 0.05 K, the presence of stacking faults between the honeycomb planes complicates its magnetic behavior. Theoretical calculations indicate that bond-dependent Kitaev interactions play a crucial role, yet Heisenberg and other non-Kitaev interactions cannot be disregarded. These systems are positioned near a tricritical point between ferromagnetic, zigzag, and incommensurate spiral orders, preventing long-range magnetic order \cite{Roser2018}. Additionally, the interlayer $O-H-O$ geometry and the absence of hydrogen ordering significantly influence their magnetic properties \cite{Yadav2018}. A related Kitaev candidate compound, \ch{Ag3LiIr2O6}, replaces hydrogen with Ag as a spacer atom. While some reports suggest Kitaev spin-liquid behavior \cite{Bahrami_T_2019}, others indicate an ordered magnetic state \cite{Bahrami2021_prb, Imai2021}. Recent studies, however, provide evidence of a low-temperature ordered phase with persistent spin dynamics \cite{AVM2021}. Magnetic order in \ch{Ag3LiIr2O6} has been confirmed through muon spin relaxation ($\mu$SR) and nuclear magnetic resonance experiments. Furthermore, anomalies in the $^{7}$Li shift and spin-lattice relaxation rate around 50 K suggest the presence of short-range correlations. A detailed $\mu$SR analysis reveals a coexistence of incommensurate N\'eel and striped magnetic environments within the ordered phase. Interestingly, significant quantum fluctuations persist deep into the ordered state, as indicated by an undiminished dynamical relaxation rate. Beyond the choice of a nonmagnetic spacer atom, modifying the in-plane magnetic ion can also lead to novel magnetic phenomena. One such system is \ch{Ag3LiRu2O6}, a $4d^4$-based compound. R. Kumar et al. \cite{RKumar2019} suggested it as an $S = 1$ system, with $\mu$SR measurements revealing spin freezing below 5 K. However, the magnetism remains at the borderline between static and dynamic even at 20 mK, as shown by longitudinal field $\mu$SR experiments. In contrast, H. Takagi et al. \cite{Takagi2022} propose that \ch{Ag3LiRu2O6} is a Mott insulator with a $J_{eff} = 0$ spin-orbit-entangled singlet state, based on spectroscopic and magnetic studies combined with quantum chemistry calculations. Beyond full substitution of either the spacer or the in-plane magnetic atom, partial substitution offers another intriguing route to tuning magnetism in honeycomb systems. For example, replacing 30\% Ru at the Ir site in $\alpha$-\ch{Li2IrO3} ($T_N = 15$ K) leads to a complete suppression of long-range order and an enhancement of Kitaev interactions \cite{Lei2014}. So, we prepared \ch{Ag3LiIr_{1.4}Ru_{0.6}O6} (will be referred to as ALIRO
henceforth) containing 30\% Ru ($4d$ atom) and 70\% Ir ($5d$
atom) on the vertices of the houeycomb to see if a suppression
of the Heisenberg interactions can be achieved
as in Ru-substituted $\alpha$-\ch{Li2IrO3}. 

Honeycomb lattices are a fertile playground for exploring Kitaev physics. The inter plane spacing can be tuned by changing the spacer atom (H, Ag, or Cu) which has been explored by us. This impacts the magnetic in-plane interactions. Also, the magnetic atoms in the honeycomb plane can be changed (Ir, Ru or a mixture of the two in our case: current paper) to tailor the magnetism of the system. The phase diagram is rich: ranging from the usual ordered state to the Kitaev Spin Liquid state with dynamic moments. Continuing our investigations in this series of systems, we focus in this paper 
on  ALIRO containing Ir$^{4+}$ ($J_{eff} = 1/2$) and Ru$^{4+}$ ($S=1$) with the  purpose of looking for possible Kitaev spin-liquid physics as an  effect of Ru-substitution. 

The remainder of the paper is organised as follows: 
we start by giving the details on sample preparation and structural characterization (section \ref{sample-preparation}) of  Ag$_{3}$LiIr$_{1.4}$Ru$_{0.6}$O$_{6}$. Following this, bulk magnetization data are presented (section \ref{sec:magnetic}). Thereafter, $^7$Li nuclear magnetic resonance (NMR) provides further validation of ordering but with additional quantum spin liquid-like features at low-$T$ in section \ref{sec:NMR}. Heat capacity (section \ref{sec:heat capacity}) does not show any signature of ordering down to 2 K and follows power-law at low-$T$. Another local probe $\mu$SR (section \ref{sec:muSR}) suggests magnetic ordering around 20 K. Finally, a comprehensive discussion and conclusions on ALIRO are presented in the sections \ref{sec:disc} and \ref{sec:conclu}, respectively. 

\section{Sample preparation and structural details}
\label{sample-preparation}

The synthesis of \ch{Ag3LiIr_{1.4}Ru_{0.6}O6} involves a two-step process similar to its parent compound \ch{Ag3LiIr2O6} \cite{Bette_C_2019,AVM2021, Bahrami2020}. In the first step, we prepared $\alpha$-\ch{Li2Ir_{0.7}Ru_{0.3}O3} using the standard solid-state reaction method\cite{Lei2014}. After obtaining phase-pure $\alpha$-\ch{Li2Ir_{0.7}Ru_{0.3}O3}, we mixed it with \ch{AgNO3} in a 1:3 molar ratio and ground the mixture. Subsequently, we heated the mixture at 270{\textdegree C} for 72 hours. In the final step, we washed the product multiple times with deionized water to remove any excess silver nitrates and by-product \ch{LiNO3}. We confirmed the absence of excess \ch{AgNO3} and by-product \ch{LiNO3} using a solution of \ch{KCl}.

To check the phase purity as well as crystal structure, we have measured the powder x-ray diffraction (XRD) of polycrystalline Ag$_{3}$LiIr$_{1.4}$Ru$_{0.6}$O$_{6}$. The XRD data were collected at room temperature with Cu-K$_{\alpha}$ radiation ($\lambda$=1.54182 \AA) over the angular range 10\textdegree $\leq$ 2$\theta$ $\leq$ 90\textdegree \space with a 0.013\textdegree \space  step size. 
\begin{figure}[!h]
	\centering
	\includegraphics[width=1.0\linewidth]{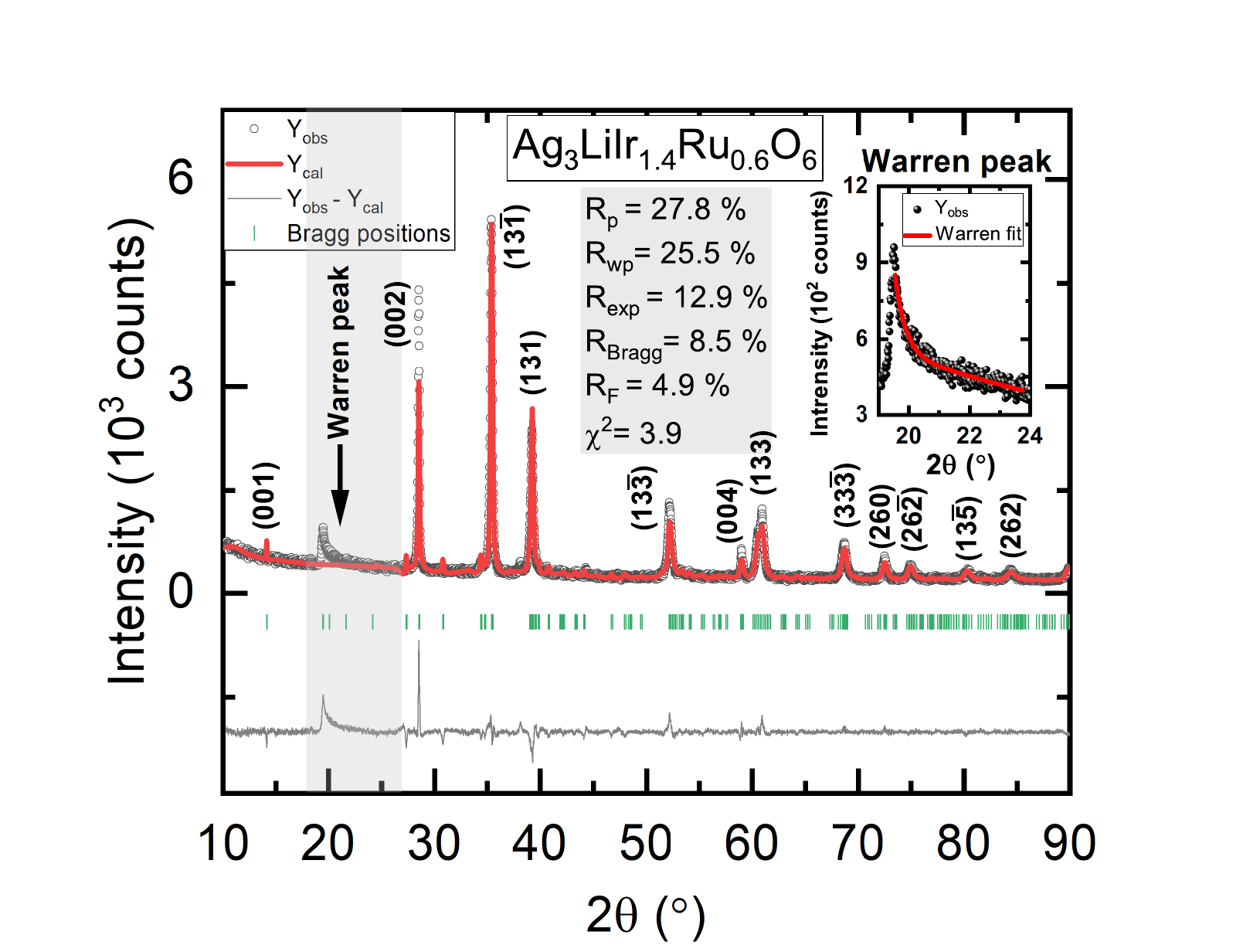}
	\caption{The single phase Rietveld refinement of XRD pattern (room temperature) of Ag$_{3}$LiIr$_{1.4}$Ru$_{0.6}$O$_{6}$. The black open circles are the observed data, the red line is theoretically calculated one, the grey line is the difference between the observed and the calculated data and the green vertical marks are Bragg positions with corresponding Miller indices. Inset shows the Warren peak).}
	\label{XRD_ALIRO}
\end{figure}
Figure \ref{XRD_ALIRO} shows the single phase Rietveld refinement of Ag$_{3}$LiIr$_{1.4}$Ru$_{0.6}$O$_{6}$. After the refinement, we found that the prepared Ag$_{3}$LiIr$_{1.4}$Ru$_{0.6}$O$_{6}$ crystallizes in the monoclinic C2/m space group. There is no sign of unreacted ingredients or impurity phase. From refinement, we obtain the cell parameters of \ch{Ag3LiIr_{1.4}Ru_{0.6}O6}, $a=$ 5.26{\AA}, $b=$ 9.11{\AA}, $c=$ 6.50{\AA}, $\alpha=\gamma=$ 90{\textdegree}, $\beta=$ 105.9{\textdegree},. The goodness of the x-ray refinement is given by the following parameters $\chi^{2}=$ 3.9, R$_{p}=$ 27.8{\%}, R$_{wp}=$ 25.5{\%}, R$_{exp}=$ 12.9{\%}. Table \ref{table_ALIRO_1} summarizes the unit cell parameters and quality factors for the Rietveld refinement of Ag$_{3}$LiIr$_{1.4}$Ru$_{0.6}$O$_{6}$. A Warren peak is observed in the range 19{\textdegree} to 24{\textdegree}, shown in inset of Figure \ref{XRD_ALIRO}. Stacking disorder was estimated to be about 7\% volume fraction. After excluding the Warren peak, Rietveld refinement is performed with faultless model (stacking faults are ignored) to extract different quality factors, atomic coordinates, site occupancies, and the isotropic Debye-Waller factors (B$_{iso}$ = 8$\pi^2$U$_{iso}$) of  Ag$_{3}$LiIr$_{1.4}$Ru$_{0.6}$O$_{6}$,which are tabulated in Table \ref{table_ALIRO_1} and Table \ref{table_ALIRO_2}, respectively.

\begin{table}[h!]
	\caption{Unit cell parameters and quality factors are reported for the Rietveld refinement of Ag$_{3}$LiIr$_{1.4}$Ru$_{0.6}$O$_{6}$ at room temperature.}
	\centering
	\scalebox{0.7}{
		\begin{tabular}{|l c| c c|}
			\hline
			Unit Cell Parameters for space group C2/m& & Quality Factors&\\
			\hline
			a(\AA) & 5.26(1) &  &  \\
			\hline
			b(\AA)& 9.11(6)& R$_{Bragg}$ (\%)&8.5 \\
			\hline
			c(\AA)& 6.50(8) & R$_F$ (\%) &4.9 \\
			\hline
			$\alpha$=$\gamma$  (\textdegree)&90 & R$_{exp}$ (\%)&12.9\\
			\hline
			$\beta$ (\textdegree)& 105.9(0) & R$_p$ (\%)&27.8\\
			\hline
			Z  & 2 & R$_{wp}$ (\%) & 25.5\\
			\hline
			V (\AA$^3$) & 300.2(6) & $\chi^2$ & 3.9\\
			\hline	
		\end{tabular}
	}
	\label{table_ALIRO_1}
\end{table}

\begin{table}[h!]
	\caption{Atomic coordinates, Normalized site occupancies, and the
		isotropic Debye-Waller factors (B$_{iso}$ = 8$\pi^2$U$_{iso}$)  are reported
		for the Rietveld refinement of Ag$_{3}$LiIr$_{1.4}$Ru$_{0.6}$O$_{6}$.}
	\centering
	\scalebox{0.7}{
		\begin{tabular}{|l| c| c| c| c| c| c|c|}
			\hline
			Atom & Wyckoff position & Site &x&y&z&Norm. Site Occ.&B$_{iso}$(\AA$^2$)\\
			\hline
			Ir(1)&4g& 2&0&0.333&0&1.4&0.4\\
			\hline
			Ru(1)&4g& 2&0&0.333&0&0.6&0.4\\
			\hline
			Li(1)&2a& 2/m&0&0&0&1&0.5\\
			\hline
			O(1)&4i&m&0.417&0&0.22&2&0.5\\
			\hline
			O(2)&8j&1&0.404&0.322&0.229&4&0.5\\
			\hline
			Ag(1)&4h&2&0&0.166&1/2&2&0.5\\
			\hline
			Ag(2)&2d&2/m&1/2&0&1/2&1&0.5\\

			\hline	
		\end{tabular}
	}
	\label{table_ALIRO_2}
\end{table}

\begin{figure}[h]
	\includegraphics[width=0.9\linewidth]{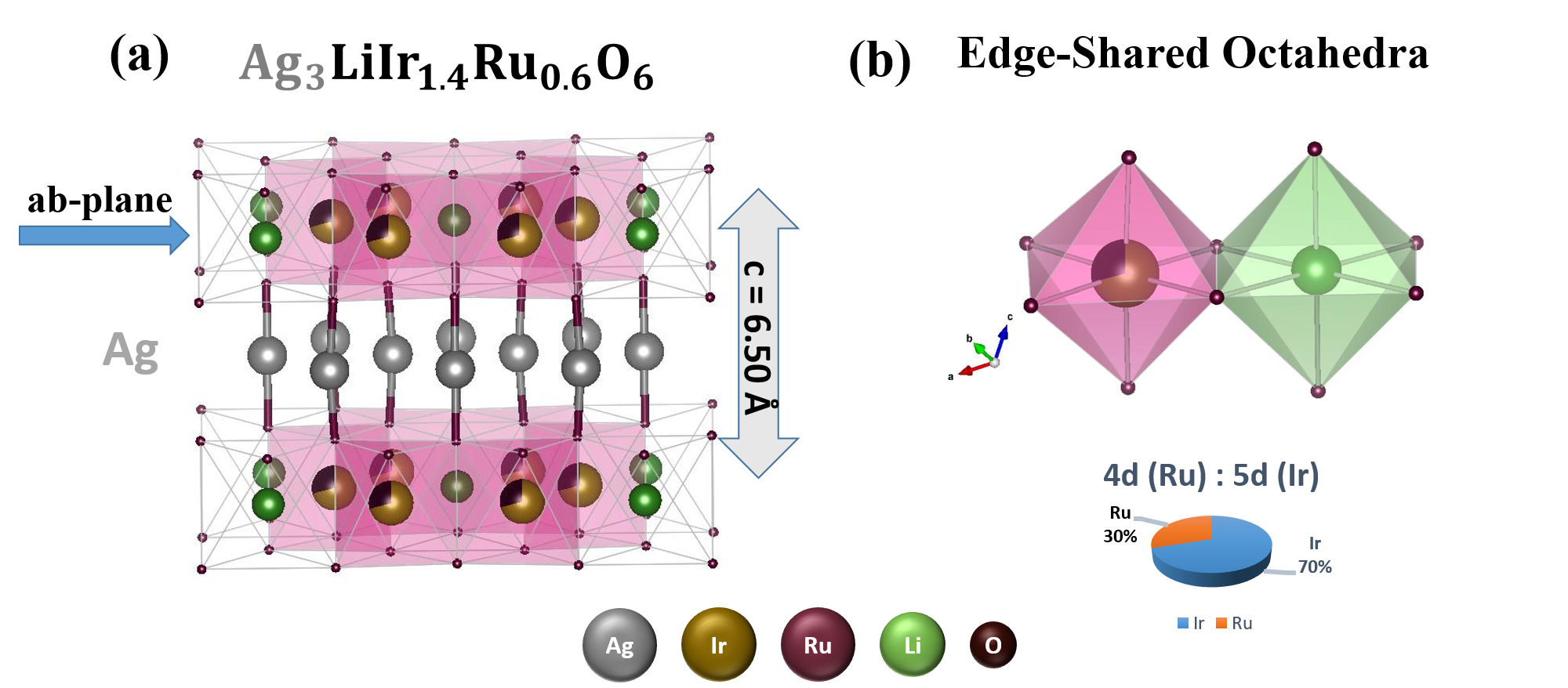}
	\caption{Crystal structure of Ag$_{3}$LiIr$_{1.4}$Ru$_{0.6}$O$_{6}$ (a) 3D view of layered material (b) Edge shared octahedra where Li and Ir/Ru sits center of Oxygen-octahedral environment. 3D pi-chart showing the Ir and Ru percentage.
}
	\label{strucFig}
\end{figure}

As illustrated in Figure \ref{strucFig}(a), the crystal structure of ALIRO has a 2D honeycomb network in the $ab$-plane, where Ir and Ru atoms occupy the vertices of the honeycombs and Li is located at the center of the honeycomb. The interlayer spacing between the honeycomb layers is $c=$6.50 {\AA}. Figure \ref{strucFig}(b) shows the edge shared octahedra where Li and Ir/Ru atoms are situated at the center of the oxygen-octahedral environment. The 3D pi-chart displays the percentage of Ir and Ru. Structural analysis reveals a nearly perfect honeycomb structure with each Ir$_{0.7}$Ru$_{0.3}$-Ir$_{0.7}$Ru$_{0.3}$ bond having a length of 3{\AA} and a bond angle of 86.9{\textdegree} for the $M-O-M$ (M = Ir$_{0.7}$Ru$_{0.3}$) bond. This is a Kitaev honeycomb geometry, where frustration arises from the bond-dependent interactions. The impact of the presence of Ir$^{4+}(5d^{5})$ and Ru$^{4+}(4d^{4})$ atoms in the honeycomb system on the magnetism is discussed in the following sections.

\section{Bulk magnetic susceptibility}
\label{sec:magnetic}

\begin{figure}[h]
	\includegraphics[width=0.9\columnwidth]{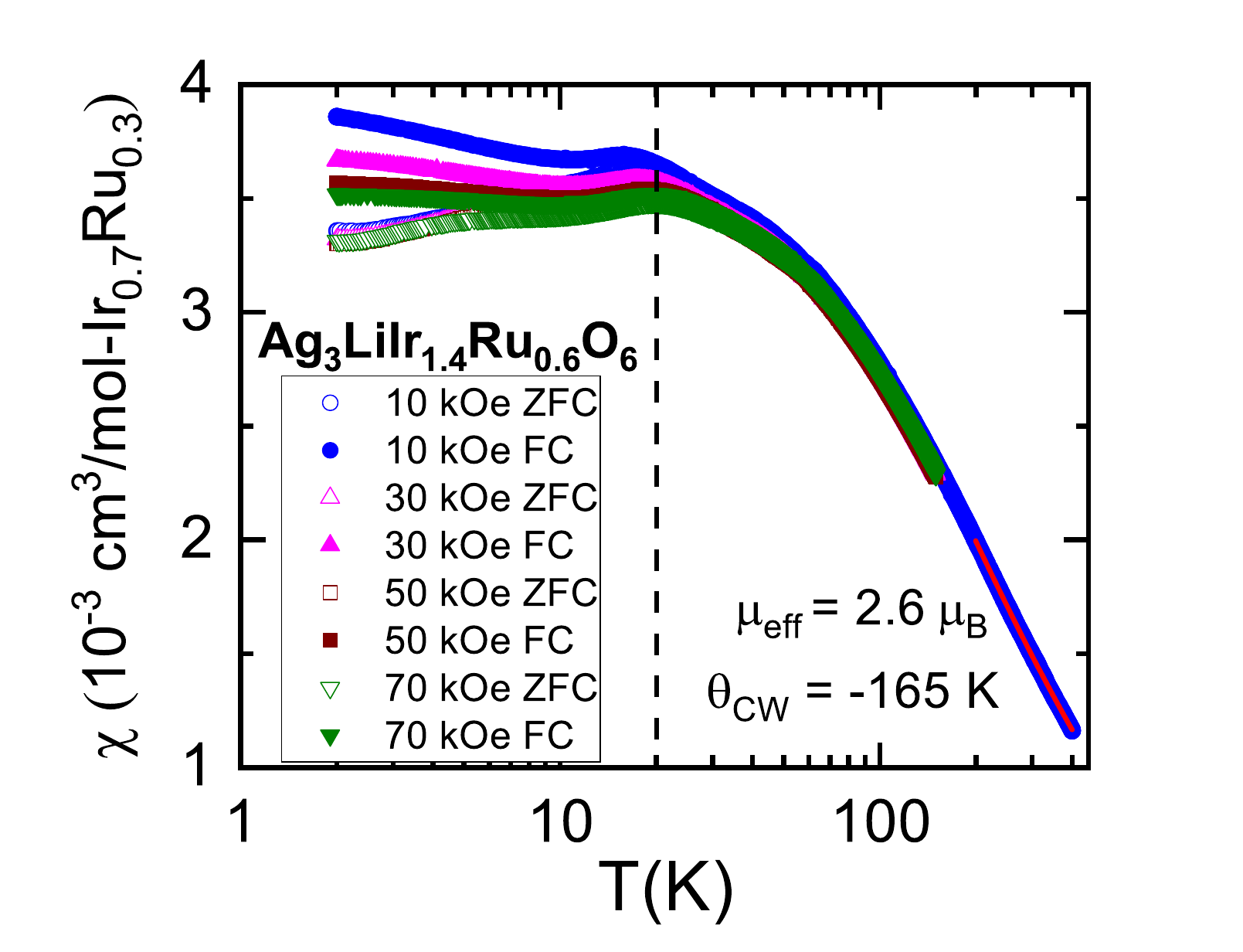}
	\caption{The temperature dependence of the susceptibility, $\chi$(T)$\equiv$$\frac{M(T)}{H}$ of Ag$_{3}$LiIr$_{1.4}$Ru$_{0.6}$O$_{6}$ measured in a 10kOe (blue), 30 kOe (pink), 50 kOe (brown), 70 kOe (green) zero field cooled (ZFC) and field cooled (FC) mode. The open circle symbol represents ZFC data and filled circles are for FC mode data.}
	\label{ALIRO_MT_10kOe}
\end{figure}

The temperature dependence of the susceptibility of ALIRO is displayed in Figure \ref{ALIRO_MT_10kOe} for different zero field cooled (ZFC) and field cooled (FC) modes in various applied fields. The plot is in a semi-log scale to emphasize the low temperature features. At around 20 K, there is a broad anomaly, possibly due to long range ordering (LRO). We performed Curie-Weiss fitting in the temperature range of 200-400 K to extract Curie constant ($C$ = 0.85 K cm$^{3}$/mol-Ir$_{0.7}$Ru$_{0.3}$) and a Curie-Weiss temperature ($\theta_{CW} \sim$ -165 K). The effective magnetic moment of \ch{Ag3LiIr_{1.4}Ru_{0.6}O6} is found to be approximately 2.6 $\mu_B$, which is larger than in Ag$_{3}$LiIr$_{2}$O$_{6}$ ($\mu_{eff}= 1.76$ $\mu_{B}$) and smaller than in Ag$_{3}$LiRu$_{2}$O$_{6}$ ($\mu_{eff}= 2.65$ $\mu_{B}$), indicating strong antiferromagnetic interaction and magnetic nature of both Ir and Ru. The ZFC-FC bifurcation at various fields suggests frozen magnetism.

\section{Nuclear magnetic resonance}
\label{sec:NMR}

In order to probe the intrinsic magnetic behavior of the correlated Ir$^{4+}$-Ru$^{4+}$ ions on the honeycomb lattice in Ag$_{3}$LiIr$_{1.4}$Ru$_{0.6}$O$_{6}$ ,  $^{7}$Li (I=3/2, $\gamma$/2$\pi$ = 16.546 MHz/T) NMR measurements were carried out.

\begin{figure}[h]
	\includegraphics[width=0.9\columnwidth]{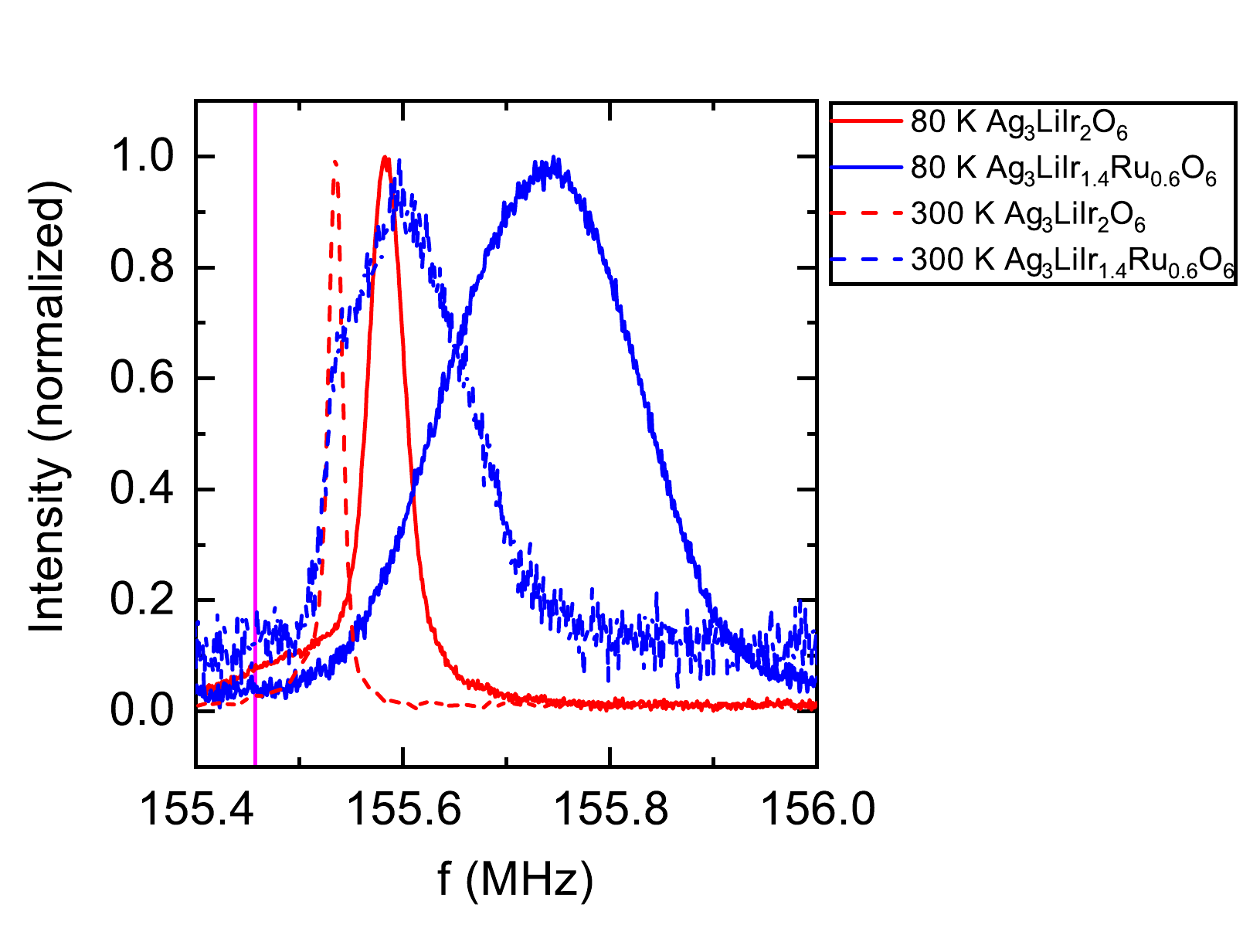}
	\caption{$^{7}$Li NMR Spectra at 300 K (dot line) and 80 K (solid line) for Ag$_{3}$LiIr$_{1.4}$Ru$_{0.6}$O$_{6}$ (blue) and Ag$_{3}$LiIr$_{2}$O$_{6}$ (red). The pink line represents reference position at 155.457 MHz.}

	\label{nmrspectra}
\end{figure}

\begin{figure}[h]
	\includegraphics[width=0.9\columnwidth]{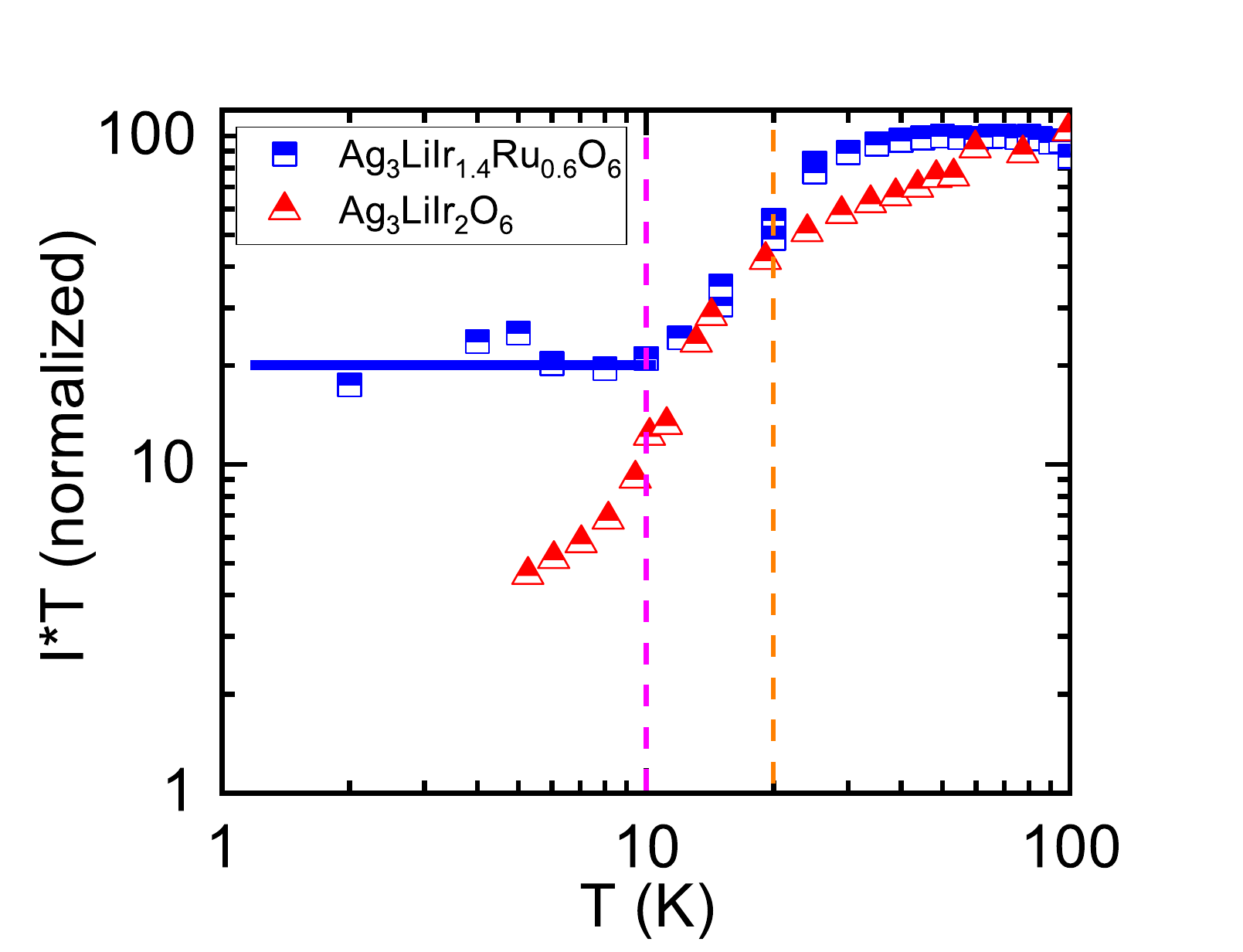}
	\caption{The product of the nuclear magnetization and temperature
		normalized to the maximum, as a function of $T$ for \ch{Ag3LiIr_{1.4}Ru_{0.6}O6} (blue square) and \ch{Ag3LiIr2O6} (red triangle).
}
	\label{normalized_intensity}
\end{figure}

Figure \ref{nmrspectra} shows representative NMR spectra at 80 K and 300 K for \ch{Ag3LiIr2O6} and \ch{Ag3LiIr_{1.4}Ru_{0.6}O6}, respectively. The black solid line at 155.457 MHz is the reference position and used for calculating the fractional shift $K$. The increase of the linewidth with decreasing temperature is connected with an increase in the internal field distribution. There is a  broadening of the NMR spectra in \ch{Ag3LiIr_{1.4}Ru_{0.6}O6} compared to \ch{Ag3LiIr2O6}. This indicates that the internal field distribution is larger in \ch{Ag3LiIr_{1.4}Ru_{0.6}O6} compared to \ch{Ag3LiIr2O6}. These are effects due to Ru-doping at Ir site in the titled honeycomb lattice. 

Figure \ref{normalized_intensity} shows the product of the nuclear magnetization $I$ and temperature $T$ normalized to the maximum, as a function of $T$ for \ch{Ag3LiIr_{1.4}Ru_{0.6}O6} and \ch{Ag3LiIr2O6}. We have observed a  rapid fall of $I*T$ at the onset of LRO at  $T\sim$ 10 K and also a clear wipe out of the  signal in case of \ch{Ag3LiIr2O6}. In \ch{Ag3LiIr_{1.4}Ru_{0.6}O6}, we observe 80{\%} of $I*T$ falls in the temperature range 10-30 K and then it remains nearly unchanged below 10 K. The fall in $I*T$ could be attributed to long range ordering although there is no clear wipe out of signal as observed in \ch{Ag3LiIr2O6}. Below 10 K, $I*T$ remains constant. That could be an  indication of the condensation of a new phase  (at least from a fraction of the sample) below  10 K.

\begin{figure}[h]
	\includegraphics[width=0.9\columnwidth]{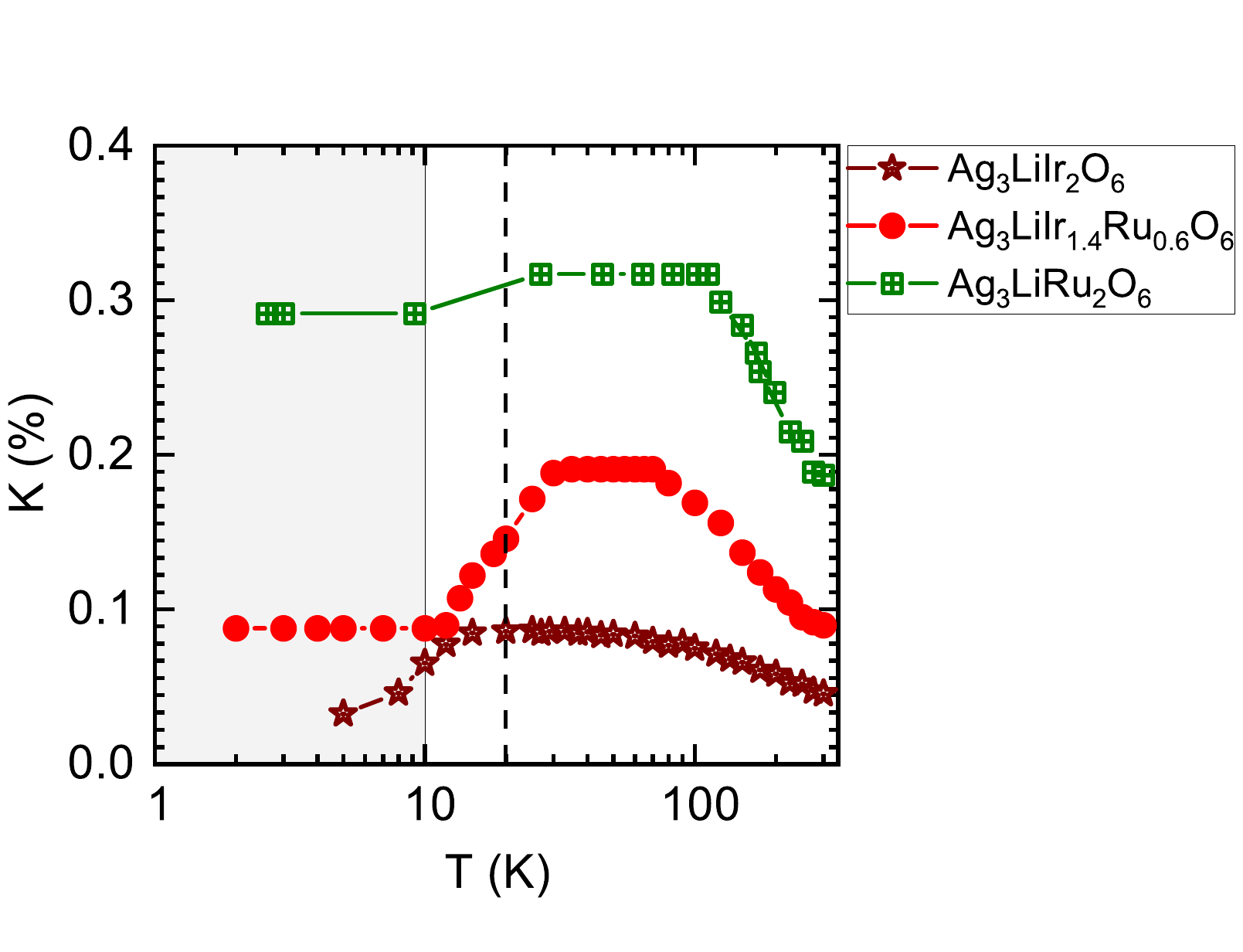}
	\caption{$^{7}$Li NMR Shift vs. T for Ag$_{3}$LiIr$_{1.4}$Ru$_{0.6}$O$_{6}$ (red) in the temperature range 300 K to 2 K. Two end compounds: Ag$_{3}$LiIr$_{2}$O$_{6}$ (brown)  and Ag$_{3}$LiRu$_{2}$O$_{6}$ \cite{RKumar2019} (olive).}
	\label{nmrshift}
\end{figure}

The NMR resonance shift probes the static intrinsic susceptibility of the system. Figure \ref{nmrshift} shows $^{7}$Li NMR shift as a function of temperature in the range 300 K to 2 K for Ag$_{3}$LiIr$_{1.4}$Ru$_{0.6}$O$_{6}$  which falls in between the two end compounds: Ag$_{3}$LiIr$_{2}$O$_{6}$  and Ag$_{3}$LiRu$_{2}$O$_{6}$.

The temperature independent behavior of $^{7}$K is one of the signatures related to quantum spin liquid-like physics. In \ch{Ag3LiIr2O6}, $^{7}$K falls below 10 K as there was magnetic ordering at $T\sim$ 10 K\cite{AVM2021}. In \ch{Ag3LiRu2O6}, $^{7}$K is temperature independent and the system was on the border line of dynamic and static spins as revealed through $\mu$SR studies. In ALIRO, $^{7}$K increases  down to 80 K in the paramagnetic region, then a $T$-independent region in the temperature range 80-30 K followed by a drop before becoming again $T$-independent in the temperature range 10-2 K. The sharp fall around 20 K could be due to LRO. The temperature independent nature of $^{7}$K below 10 K has some signatures of a gapless ground state.

\begin{figure}[h]
	\includegraphics[width=1.0\columnwidth]{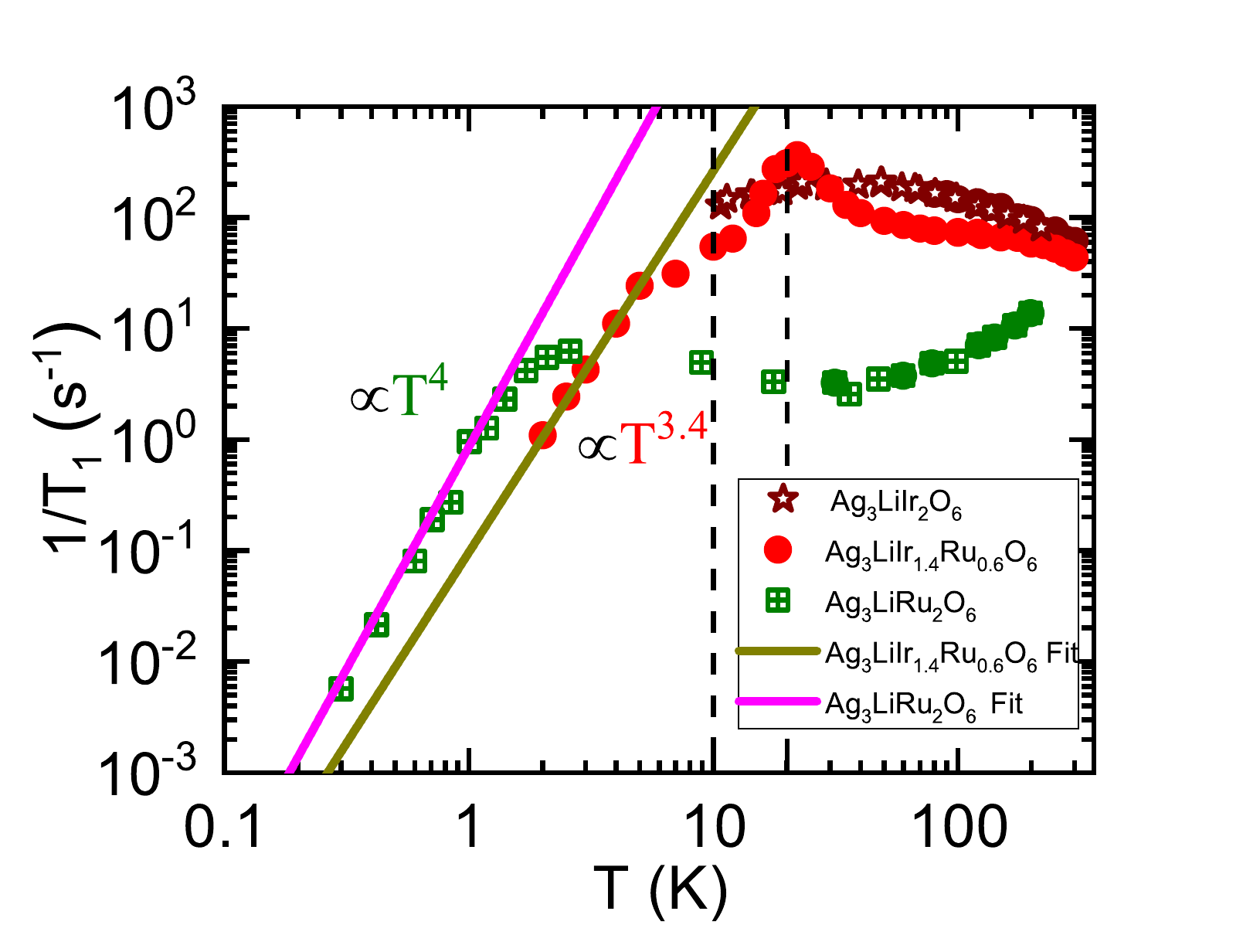}
	\caption{The Variation of 1/$T_1$ as a function of temperature for Ag$_{3}$LiIr$_{1.4}$Ru$_{0.6}$O$_{6}$ (red) in the temperature range 300 K to 2 K. Two end compounds: Ag$_{3}$LiIr$_{2}$O$_{6}$ (brown)  and Ag$_{3}$LiRu$_{2}$O$_{6}$ \cite{RKumar2019} (olive).}
	\label{spinlattice}
\end{figure}

The spin-lattice relaxation rate, (1/$T_{1}$) measures the $q$-averaged imaginary part of susceptibility, $\chi"(q,\omega_{0})/\omega_{0}$ at a very low-energy $\omega_{0}$, which reflects the density of low-energy spin excitations. We measured $^{7}$Li spin-lattice relaxation rate (1/$T_{1}$) for \ch{Ag3LiIr_{1.4}Ru_{0.6}O6} by using a saturation recovery method for various temperatures from 300 K to 2 K.

The Figure \ref{spinlattice} shows the spin-lattice relaxation rate, 1/$T_{1}$ as a function of temperature for Ag$_{3}$LiIr$_{1.4}$Ru$_{0.6}$O$_{6}$  and two end compounds: Ag$_{3}$LiIr$_{2}$O$_{6}$   and Ag$_{3}$LiRu$_{2}$O$_{6}$. A broad anomaly at  $T\sim$ 50 K was observed in \ch{Ag3LiIr2O6} indicating of short range ordering followed by a wipe out of the signal below 10 K due to LRO \cite{AVM2021}. In \ch{Ag3LiRu2O6}, a power-law variation in the spin-lattice relaxation rate indicates gapless excitations. The temperature variation of the spin-lattice relaxation rate, 1/$T_{1}$ in \ch{Ag3LiIr_{1.4}Ru_{0.6}O6} has a broad anomaly at $T\sim$ 20 K followed by a power-law variation at low-$T$. That indicates long range ordering at $T\sim$ 20 K followed by gapless excitations at low-$T$ through the condensation of a  new phase.

\section{Muon spin relaxation}
\label{sec:muSR}

\begin{figure}[h]
	\includegraphics[width=1.0\columnwidth]{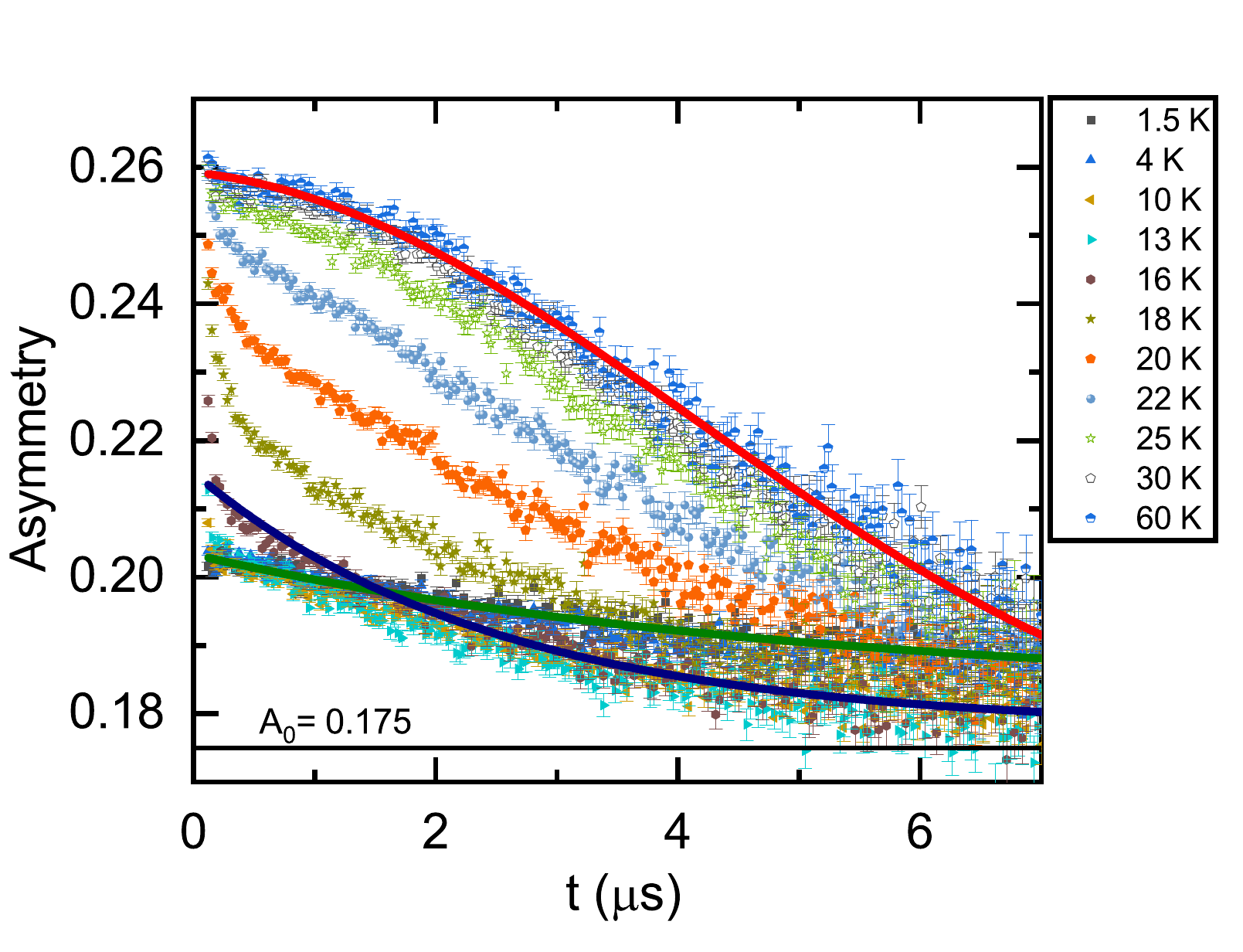}
	\caption{Variation of the muon asymmetry with time is shown at selected
		temperatures. Clear 2/3-loss in the initial asymmetry is seen below 18 K
		indicative of long-range magnetic order. Fits at some representative
		temperatures are shown as explained in the text.} 
	\label{musr-asymmetry}
\end{figure}

We have measured zero field (ZF) muon asymmetry in the temperature range 1.5 K to
60 K (Figure \ref{musr-asymmetry}). We observe a 2/3-loss in the initial asymmetry at the lowest temperature (1.5 K). The high
oscillating frequency of muon spins due to static local moments leads to a loss of 2/3-
component of initial asymmetry. The fast-falling component could not be capture in the ISIS MUSR
instrument which has a pulsed muon source. 

\begin{figure}[h]
	\includegraphics[width=1.0\columnwidth]{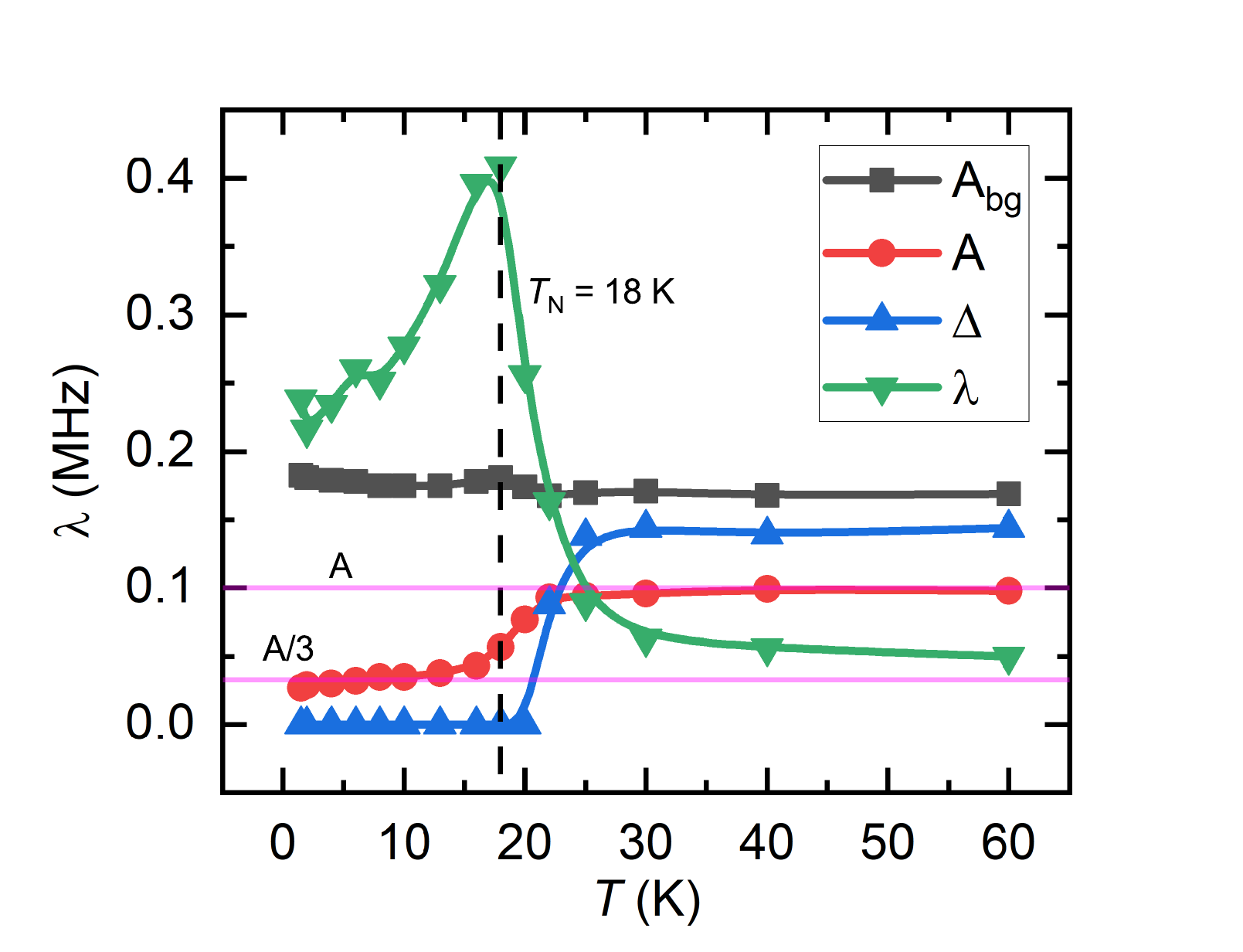}
	\caption{Variation of the muon relaxation rate, $\lambda$ as a function of temperature
		for ALIRO (Green triangle). Static nuclear field distribution, $\Delta$ is about 1.8 Oe
		and its variation with T is shown with blue triangle. Relaxing asymmetry, A
		with T is shown with red circle. A is constant and no loss down to 25 K. Below
		25 K, a fast and a slow component has evolved with lowering temperature.
		Below 13 K, due to large static field in ordered state, fast component could not
		observe. The slow component turns out to be 1/3-component of initial
		asymmetry. $A_{bg}$ is nearly temperature independent (black square).}
	\label{musr-lambda}
\end{figure}

Remaining 1/3-component of initial asymmetry decays exponentially. Such a loss of initial asymmetry
often seen in magnetically ordered system due to large static spin in ordered state. We
find that at temperatures above 20 K, the data are well fit to a product of a static Kubo-
Toyabe function with an exponential in addition to a constant background $A_{0}$ i.e.,
$A(t)= A_{rel}G_{KT}(\Delta, \textit{T}) exp(- \lambda t) + A_{0}$. Here, $G_{KT} (\Delta, \textit{T})$ is the Kubo-Toyabe function which models the relaxation of muons in a Gaussian distribution of magnetic fields from nuclear moments and $A_{rel}$ is relaxing asymmetry. From these fits we obtain the
field distribution $\Delta$ to be about 1.8 Oe. This value is typical of nuclear dipolar fields at the muon site, in the present case arising from $^{107,109}$Ag, $^{6,7}$Li, $^{99,101}$Ru and $^{191,193}$Ir nuclei. The exponential term $exp(- \lambda t)$ arises from the relaxation due to fluctuations of
the electronic local moments. Below 20 K, muon asymmetry was well fitted with
single exponential in addition to a constant background $A_{0}$ i.e., $A(t)= A_{rel} exp(- \lambda t) + A_{0}$. This relaxation rate $\lambda$(T) is small at high temperatures and gradually increases as the local moment fluctuation rate gets smaller (Figure \ref{musr-lambda}). We notice a sharper increase of $\lambda$(T) below about 18 K. $\lambda$(T) decreases with temperature below 18 K and down to 1.5 K. Below ordering temperature, presence of static spins slow down the muon spin relaxation rate, $\lambda$.

\section{Heat capacity}
\label{sec:heat capacity}

\begin{figure}[h]
	\includegraphics[width=1.0\columnwidth]{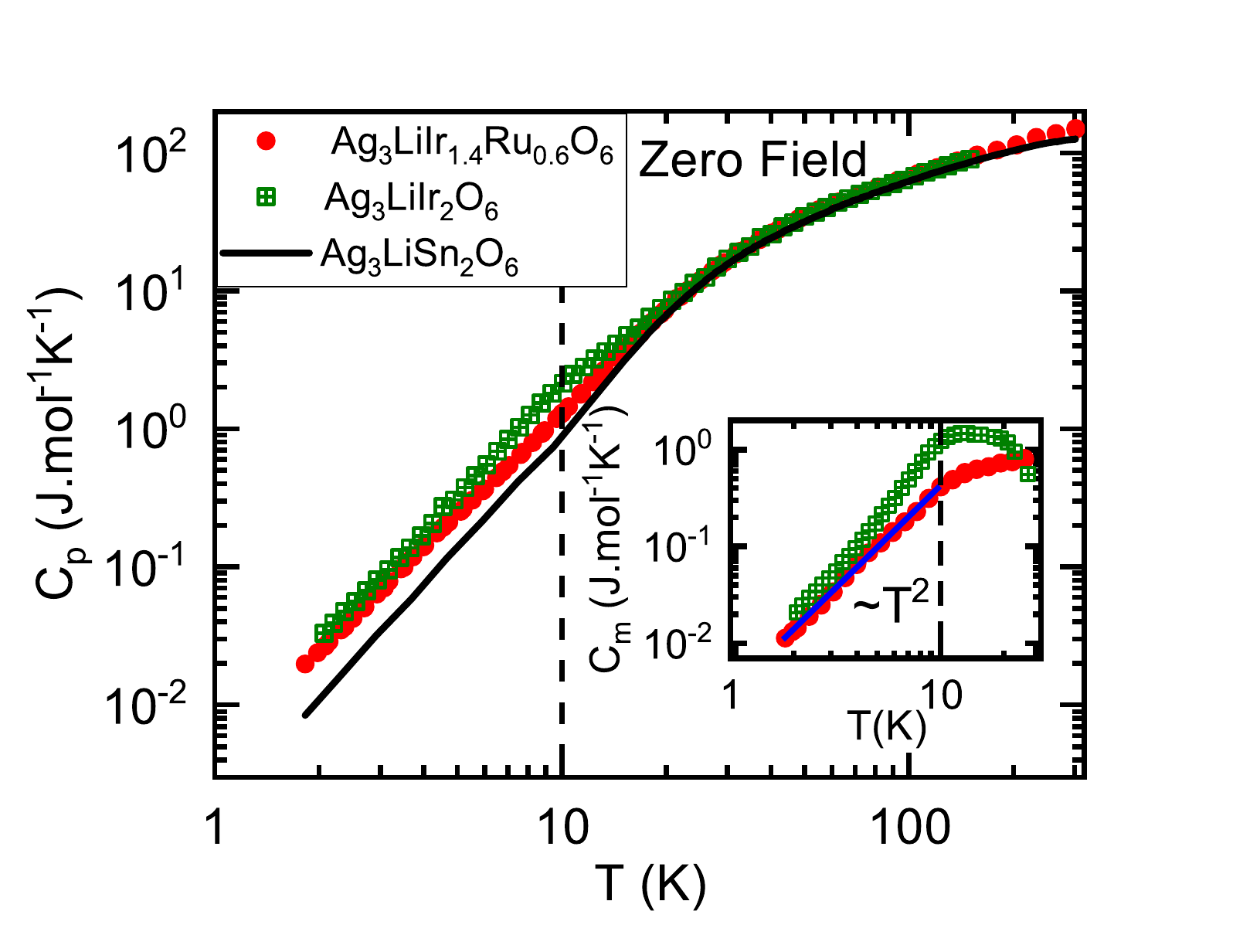}
	\caption{Specific heat, C$_{p}$ as a function of temperature for Ag$_{3}$LiIr$_{1.4}$Ru$_{0.6}$O$_{6}$ (red circle), Ag$_{3}$LiIr$_{2}$O$_{6}$ (olive square cage) and non-magnetic analog Ag$_{3}$LiSn$_{2}$O$_{6}$ (black line). (Inset) Magnetic heat capacity comparison between Ag$_{3}$LiIr$_{1.4}$Ru$_{0.6}$O$_{6}$ (red circle) and Ag$_{3}$LiIr$_{2}$O$_{6}$ (olive square cage).}
	\label{HeatCapacity}
\end{figure}

To get more information about low energy excitations and ordered phase in ALIRO, we measured the heat capacity of the sample at constant pressure $C_{p}(T)$ at different fields (0-70 kOe) in the temperature range (2-300 K) for 0 kOe and (2-40 K) for other fields. There is no field dependency in heat capacity down to 2 K. Figure \ref{HeatCapacity} shows zero field specific heat, C$_{p}(T)$ as a function of temperature for Ag$_{3}$LiIr$_{1.4}$Ru$_{0.6}$O$_{6}$, Ag$_{3}$LiIr$_{2}$O$_{6}$  and non-magnetic analog Ag$_{3}$LiSn$_{2}$O$_{6}$. 

We have created a log-log plot to more clearly show the low-temperature region. In \ch{Ag3LiIr2O6}, there was a hint of an  ordering peak at $T\sim$ 10 K although the peak is not quite sharp perhaps because of the onset of short range ordering already at $T\sim$ 50 K\cite{AVM2021}. There is a clear suppression of the 10 K ordering peak (of Ag$_{3}$LiIr$_{2}$O$_{6}$) in Ag$_{3}$LiIr$_{1.4}$Ru$_{0.6}$O$_{6}$. In zero-field, the difference between the  measured $C_{p}$ of the magnetic compound and the $C_{p}$ of the non-magnetic analog (that is the lattice contribution) is the magnetic specific heat, $C_{m}$ of the compound. The specific heat for \ch{Ag3LiIr2O6}, \ch{Ag3LiIr_{1.4}Ru_{0.6}O6} and its non-magnetic analog, \ch{Ag3LiSn2O6}  match above $\sim$ 30 K. This indicates that the lattice contribution dominates at high temperature. Below 30 K, there are clear differences between $C_{p}$ of \ch{Ag3LiIr2O6}, \ch{Ag3LiIr_{1.4}Ru_{0.6}O6} and non-magnetic analog, \ch{Ag3LiSn2O6}. The calculated $C_{m}$ for \ch{Ag3LiIr2O6} and \ch{Ag3LiIr_{1.4}Ru_{0.6}O6} are plotted in a log-log scale and shown in inset of Figure \ref{HeatCapacity}. In the inferred $C_{m}$ vs $T$ of \ch{Ag3LiIr2O6}, there is a clear ordering peak around 10 K (more prominent than in the $C_{p}$ vs $T$ plot) and then it follows an exponential decrease. In \ch{Ag3LiIr_{1.4}Ru_{0.6}O6}, it is not clear whether the weak anomaly in $C_{m}$ vs $T$ in the range 10-20 K is associated with any frozen magnetism or long range ordering. But, a power law ($T^{2}$) variation at low-$T$ (below 10 K) in the magnetic heat capacity of Ag$_{3}$LiIr$_{1.4}$Ru$_{0.6}$O$_{6}$ is observed. This indicates gapless excitations. We speculate that this arises from a new phase which condenses (at least from a fraction of the sample) at  low-$T$ \cite{Baenitz2017}.

\section{Discussion}
\label{sec:disc}

From the extensive range of measurements we have presented, which includes both bulk and local probes, let us now take a look from a broader perspective.
We set out with the motivation of suppressing the Heisenberg interaction in \ch{Ag3LiIr2O6}, aiming to enhance the anisotropic bond dependent interactions that could promote a Kitaev quantum spin liquid state. Our approach was inspired by previous work on \ch{Li2Ir_{1-x}Ru_{x}O3} \cite{Lei2014}, where it was demonstrated that substituting 30\% Ru in place of Ir could effectively tune the exchange interactions to suppress the ordering. Encouraged by these findings, we pursued the intercalation of \ch{Li2Ir_{0.7}Ru_{0.3}O3} with Ag to further explore the effect of structural and electronic modifications on the magnetic properties.
Our intercalation experiments were successful, leading to the stabilization of \ch{Ag3LiIr_{1.4}Ru_{0.6}O6}. X-ray diffraction and Rietveld refinement suggests that a single phase pure compound has been formed which crystallizes in the  C2/m space group like its parent compound, \ch{Ag3LiIr2O6}. The Curie-Weiss temperature was inferred to be $\sim$ -165 K and the effective magnetic moment is found to be approximately 2.6 $\mu_{B}$, indicating strong antiferromagnetic interactions and magnetic nature of both Ir and Ru. Presence of magnetic ordering below approximately 20 K in ALIRO was revealed by bulk magnetization, loss in the NMR signal intensity (though not a  complete wipe out like in \ch{Ag3LiIr2O6}), anomaly in the $^7$Li spin lattice relaxation rate, 1/$T_{1}$ and the muon spin relaxation rate, $\lambda$. A 2/3$^{rd}$-loss in the initial zero field muon asymmetry at the lowest temperature (1.5 K), is indicative of a static spin-ordered state. Although ALIRO is magnetically ordered (an \textbf{upward} shift in the ordering temperature is an effect of the 30\% Ru-substitution at the Ir site in \ch{Ag3LiIr2O6}) like its parent compound, \ch{Ag3LiIr2O6} but low-$T$ features are significantly in contrast. Some curious features do show up below 10 K which suggest that a new phase might have been stabilized. In heat capacity, it is not clear whether there is any weak anomaly around 20 K but a $T^{2}$ power-law behavior has been observed below 10 K. There is no sign of a complete wipeout of the NMR signal intensity even down to 2 K unlike in  \ch{Ag3LiIr2O6}. Together with this, a $T$-independent NMR shift and a power law variation of $1/T_1$ are observed below 10 K. These features might suggest condensation of a new spin liquid like phase (at least from a fraction of the sample) at low-$T$. It is possible that  two competitive exchange energies due to the two types of magnetic ions (Ir and Ru) leads to two magnetic phases (long range ordering $\sim$ 20 K and gapless excitations at low-$T$ through condensation of new spin liquid like phase) in different temperature range.

\section{Conclusions}
\label{sec:conclu}

Ag$_{3}$LiIr$_{1.4}$Ru$_{0.6}$O$_{6}$, a Kitaev honeycomb quantum material containing magnetic Ir$^{4+}$ and Ru$^{4+}$ shows spin liquid like features at low-$T$ (below 10 K which is below the 20 K ordering temperature) coming from about 20\% of non-freezing moments. Muon spin relaxation, nuclear magnetic resonance and magnetization confirms the shift of the ordering temperature to 20 K in Ag$_{3}$LiIr$_{1.4}$Ru$_{0.6}$O$_{6}$ from 10 K  of \ch{Ag3LiIr2O6}. Heat capacity, $^7Li$-NMR shift and NMR-$1/T_{1}$ suggests condensation of a spin liquid-like phase at low-$T$ with gapless excitations. Nonetheless, data presented here suggest Ru-substituted \ch{Ag3LiIr2O6} to be a strong Kitaev quantum Spin liquid candidate. A single crystal study on this compound could be part of future work.

\section{acknowledgment}

We thank MOE India, STARS project ID:358 for
financial support. We thank Central Facilities at
IIT Bombay for support for various measurements.
Experiments at the ISIS Neutron and Muon Source
were supported by a beam-time allocation RB2220486
from the Science and Technology Facilities Council
(https://doi.org/10.5286/ISIS.E.RB2220486).

\bibliographystyle{apsrev}
\bibliography{ref}

\end{document}